\documentclass[transmag]{IEEEtran}
\usepackage{latexsym}
\usepackage{graphicx}
\usepackage{amsfonts,amssymb,amsmath}
\usepackage{hyperref}
\usepackage{float}
\usepackage{authblk}
\def\BibTeX{{\rm B\kern-.05em{\sc i\kern-.025em b}\kern-.08em T\kern-.1667em\lower.7ex\hbox{E}\kern-.125emX}}
\begin{document}

\title{Portfolio Management using Deep Reinforcement Learning}

\author[1]{Ashish Anil Pawar}
\author[2]{Vishnureddy Prashant Muskawar}
\author[3]{Ritesh Tiku}
\affil[1]{pawaraa16.it@coep.ac.in}
\affil[2]{muskawarvp16.comp@coep.ac.in}
\affil[3]{ritesh16.comp@coep.ac.in}

\IEEEtitleabstractindextext{\begin{abstract}Algorithmic trading or Financial robots have been conquering the stock markets with their ability to fathom complex statistical trading strategies. But with the recent development of deep learning technologies, these strategies are becoming impotent. The DQN and A2C models have previously outperformed eminent humans in game-playing and robotics. In our work, we propose a reinforced portfolio manager offering assistance in the allocation of weights to assets. The environment proffers the manager the freedom to go long and even short on the assets. The weight allocation advisements are restricted to the choice of portfolio assets and tested empirically to knock benchmark indices. The manager performs financial transactions in a postulated liquid market without any transaction charges. This work provides the conclusion that the proposed portfolio manager with actions centered on weight allocations can surpass the risk-adjusted returns of conventional portfolio managers.\end{abstract}

}

\maketitle

\section{INTRODUCTION}

\IEEEPARstart{A}{lgorithmic} trading involves periodic decision-making for allocation of the amount of investment to each asset in the portfolio. Conventionally decision-making making has been based on a blend of statistical tools and human-emotions. These algorithms were a set of complex conditional rules delivering outstanding returns to the portfolio in a previously simulated state. But they underperformed in an unwonted hostile state. The recent advancement in the area of deep reinforcement learning had made it plausible for the algorithm to take into account a wide variety of states. Numerous studies have been published to equip the algorithm in these hostile states. The studies had an agent and an environment to interact with each other to compose better rewards. The agent here could take three actions to buy, hold, or sell at a time step and the environment would reward it accordingly. Over time, the agent would learn to act with an incentive to get high returns on the asset. The learning was then extended to the portfolio management task with a set of three actions for each asset in the portfolio. In our proposed work, the agent is tasked to allocate weights to each asset and the reward is measured in the returns gained from the allocation. 

The weight alteration issues numerous financial transactions requiring buying and selling assets in financial markets. Transactional orders are placed on each working day amounting to 252 days annually. The financial markets can be very unpredictable so we consider a postulated market very our every order is fulfilled which may not be always the case in actual markets.

\section{Literature Survey}
\subsection{Financial Portfolio Management}
\subsubsection*{Introduction}
In \cite{ref6} Portfolio management, capital is continuously re-allotted in different financial entities.At constant time periods, one needs to do some decisions and actions regarding investment for a trading bot.
\subsubsection*{Method}
The model initiates the portfolio weights vector by allocating high weights to 11 most volumed non-cash assets and bitcoin. After a time "t" a tensor is provided having rank 3, having the number of assets , excluding cash, the number of periods before "t", and the number of features which is equal to 3 here. The policy network, called the Ensemble of Identical Independent Evaluators (EIIE), takes a 3D tensor of historical market-relative price data as input, as well as the vector of portfolio weights from the previous period. The output of the EIIE is the portfolio weight vector for the subsequent period. The determinist policy gradient is used by gradient ascension to achieve an optimal strategy.
\begin{figure}[H]
    \centering
    \includegraphics[scale=0.20]{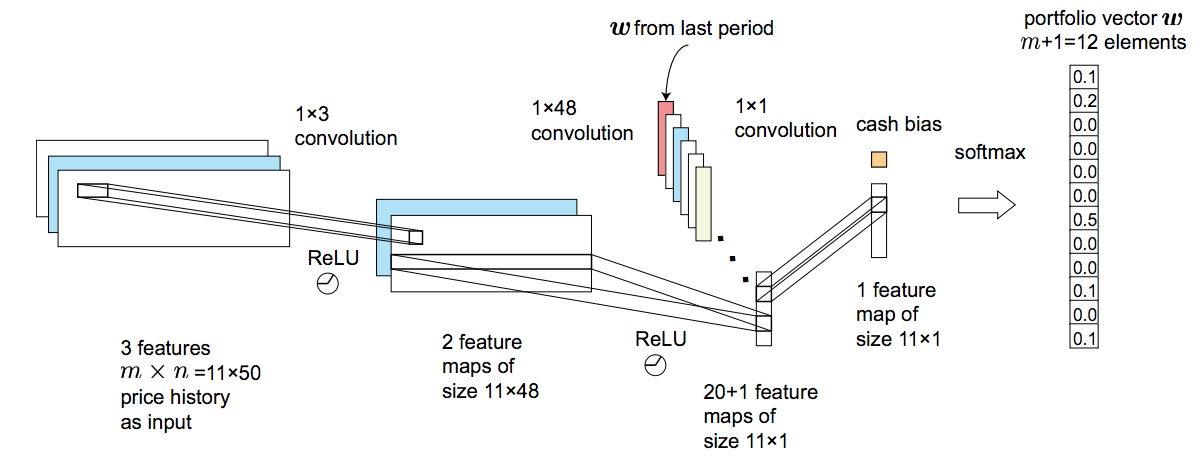}
    \caption{Portfolio Management System}
    \label{fig:galaxy}
\end{figure}

The EIIE is trained at the end of each trading period on stochastic mini-batches of historical data. Mini-batch intervals are sampled from a geometric distribution so recent data is selected more often than older data. The agent maintains a memory of the portfolio weight vector at each trading period, known as the Portfolio Vector Memory (PVM). The PVM is overwritten both when the agent is redistributing the portfolio and during training. For each mini-batch, the agent ascends the reward gradient of the interval by an amount determined by the learning rate. 

The performance of portfolio weights from the previous trading period is inserted into the networks. To minimise transaction costs, the portfolio management agent needs to prevent itself from significant changes between consecutive portfolio vectors. This is accomplished by implementing specially LSTMs Recurrent subnets which-it from the weights exploding.

\subsubsection*{Performance Evaluation}
The profitability of the model exceeds all surveyed conventional portfolio selection approaches, as shown in the paper by the results of three back-test studies in a crypto-currency market over different times. The structure was realised in these experiments using three distinct networks(CNN, RNN, LSTM). all three models performed better than other trading algorithms in the final cumulative portfolio value. Also in all three tests the EIIE networks monopolised the top three positions in the risk-adjusted ranking, indicating the consistency of the system in its results.

\subsubsection*{Shortcomings}
\begin{enumerate}
  \item The portfolio allocation needs to be executed at high frequency to avoid the stochastic noise of the market.
  \item The liquidity of the market needs to be contemplated.
\end{enumerate}

\subsection{Deep Convolution Neural Network Approach}

\subsubsection*{Introduction}
A novel CNN-TA model of algorithmic trading that uses a 2-D convolutionary neural network based on image recognition is described in \cite{ref7}.

\subsubsection*{Method}
Each day a 15x15 pixel image data is produced using 15 technical indicators (EMO, PPO, etc) and 15 different periods. The image data is then marked with a naive algorithm that classifies the bottom points, top points, and middle points(buy, sell, hold) The CNN is then equipped with the following functions:-
\begin{enumerate}
    \item The input layer (15x15)
    \item Two convolution layers (32 and 64, 3x3 kernels respectively)
    \item Max pooling layer (2x2)
    \item Two dropout layers (0.25, 0.50)
    \item Fully connected layer (128)
    \item Output layer (3)
\end{enumerate}

\subsubsection*{Performance Evaluation}
The proposed network was evaluated on ETFs and DOW30 over an interval of 5 years from 2012 to 2017 and tested in the year 2018. The network gave an average annualized return of 13.01 percent, performing significantly better than conventional statistical models.

\subsubsection*{Shortcomings}
\begin{enumerate}
    \item The model studied a long-only strategy but applying a long-term short strategy could dramatically increase the profit as there are many idle periods when the model can use liquid assets but does not as it is waiting for a signal to buy.
    \item A portfolio of multiple stocks / ETFs can be dynamically distributed based on the stock / ETF performance and improved overall performance can be achieved with less risk.
    \item Lack of analyzing the correlation between the indicators to create more insightful images for the model.
    \item Hyperparameters like image size, depth of the network, etc were not tuned.
    \item The technical indicators optimization could have produced better results.
\end{enumerate}

\subsection{Deep Reinforcement Learning Approach}

\subsubsection*{Introduction}
Explore the Deep Deterministic Policy Gradient (DDPG) deep reinforcement learning algorithm to find the best trading strategy on the complex and competitive stock market methodology is used in \cite{ref5}.

\subsubsection*{Method}
If we take the process to be an MDP, we can assume the problem to be of maximising nature.
\begin{itemize}
    \item State s = [p; h; b] - Where p, h and b are, stock price information,  number of stock holdings and  balance remaining respectively.
    \item Action a - Contains all the sell, buy and hold related actions on a stock as a set.
    \item Reward r(s; a; s1) - Reward can be positive or negative based on the change in value of portfolio.It is measured as the sum of the shares of all stocks holding p T h and the balance b.
    \item Policy (s) - the trading strategy of stocks at state s. It is essentially the probability distribution of a at state s. 
    \item Action-value function Q(s; a) - the expected reward achieved by action a at state s following policy.
 \end{itemize}
 
\begin{figure}[H]
    \centering
    \includegraphics[scale=0.1]{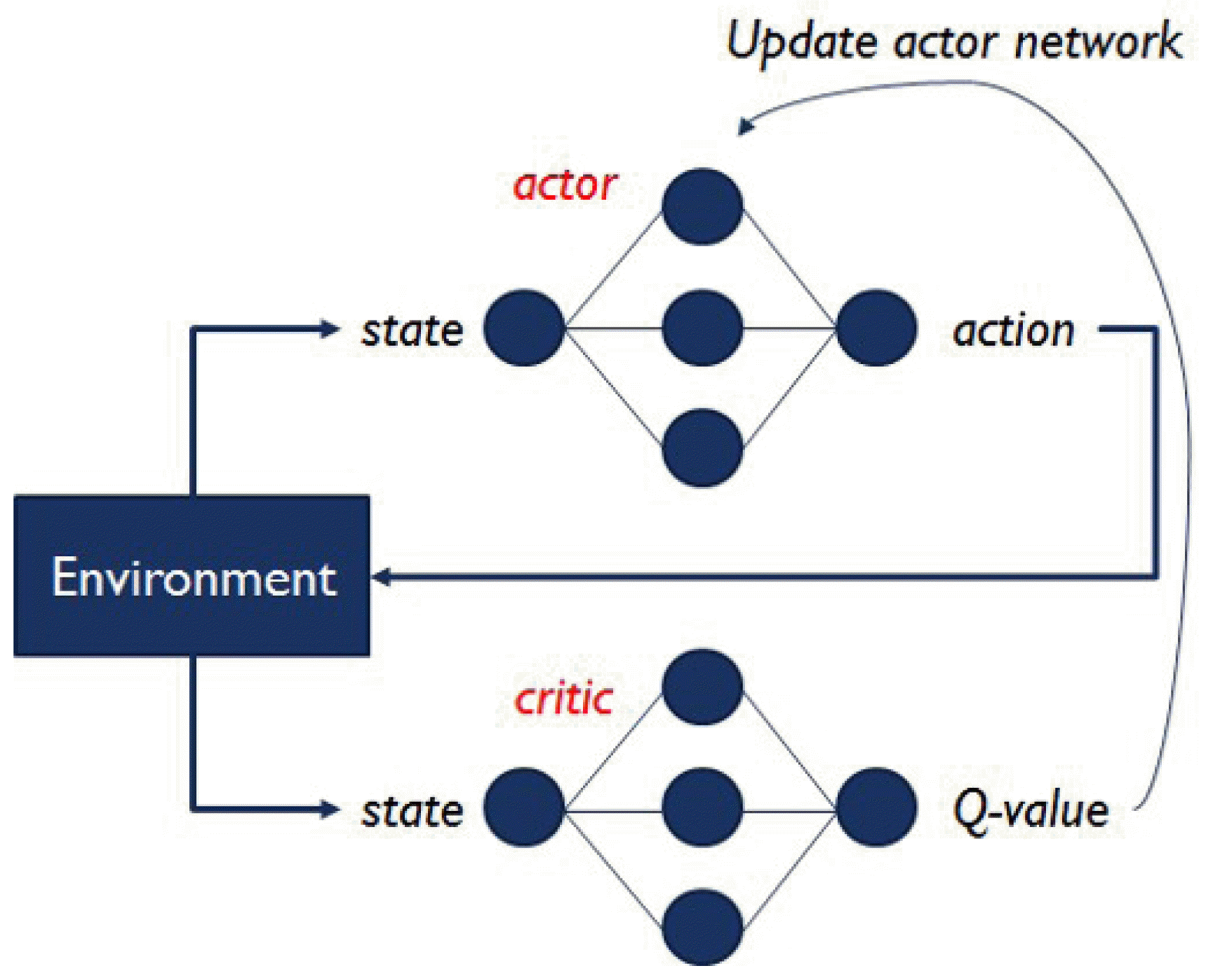}
    \caption{A typical DDPG Flow Chart}
    \label{fig:galaxy}
\end{figure}

The DDPG boils down the problem to optimize the policy that maximizes the action-value function at a terminal time. Utilises an actor critic system where actor maps states to actions and critic rewards the action under that state. Noise is added for overfitting hindrance.

\subsubsection*{Performance Evaluation}
The dataset is partitioned into a training set that covers the 2009-2015 period, the 2016-2017 validation set and the 2018 stock data test set, respectively. The model ran on an environment of 30 stocks daily traded on NYSE. It outperformed the Dow Jones Industrial Average by lucratively producing a 25.87 percent annualized return on the test data.

\subsubsection*{Shortcomings}
\begin{enumerate}
    \item The model performs well on low scale data but underperforms when given more than large scale data.
    \item The decision making has high latency which makes it incompatible with high-frequency trading architectures.
    \item The Sharpe ratio can be improved if we take into account risk-free interest rates while calculating rewards.
\end{enumerate}

\subsection{Deep Attention Recurrent Q-Learning}

\subsubsection*{Introduction}
Paper \cite{ref1} explains why DARQN is better than DQN. DQN decides on the next optimal action based on the visual information which corresponds to the agent's 4 last game states. Thus, the algorithm can not master certain games that in the past involve a player to recall events that are more distant than four screens. This is for the case of Atari 2600 games.

\subsubsection*{Method}
The DARQN architecture consists of convolutional, Attention, and recurrent networks. 
\begin{enumerate}
    \item CNN receives a representation of the current game state st in the form of a visual frame at each time phase t, on the basis of which it generates a set of D feature maps, each of which has a dimension of m*m. 
    \item The attention network transforms these maps into a set of vectors vt = \{v1 t,...,vL t \}, vi t $\varepsilon$ RD, L = m * m and outputs their linear combination zt  $\varepsilon$  RD, called a context vector.
    \item  The LSTM in this case takes an input comprising of a context vector, previous hidden state $ht-1$ and memory state $ct-1$. The output is a hidden state ht that is used to evaluate Q-value of each action that the agent can take in state st, and for generating the attention network to find a context vector at t + 1.
\end{enumerate}

Each vector vi at t comprises of features extracted by CNN in different regions of the screen, and the context vector is a weighted avg of all vi in DARQN.
In compliance with any stochastic strategy of attention $\pi$g, the "strong" attention method involves sampling only one attention position from L available at each time stage t.
The given algorithm was tested on several popular Atari 2600 games: Breakout, Donkey Kong, Space Invaders,Pac-man, asteroids.

\subsubsection*{Results}
\begin{enumerate}
    \item Both DARQN models show high efficiency. But the hard-attention-based agent tends to be inferior to the soft one.
    \item LSTM versions do poorly than the original DQN. The low number of unrolling steps used when training the LSTM network is one potential explanation for that.  
\end{enumerate}

\subsubsection*{Shortcomings}
Q-learning Algorithm, although popular, can overestimate action values under certain conditions. This hampers the performance in some cases.

\subsection{Double Q-Learning}
\subsubsection*{Introduction}
The research paper \cite{ref4} shows the effects of overestimations caused by DQN. To prevent the overestimation caused by the Q-Learning algorithm, the research paper uses a double Q-Learning algorithm. It also claims to increase the performance of the adapted DQN.

\subsubsection*{Method}
\begin{enumerate}
    \item In Q-Learning and DQN, the value function to select an action, as well as the DQN, both use the same value of parameters. This is what majorly leads to overestimation.
    \item Hence, the selection is decoupled from the evaluation in Double Q-Learning.
    \item Initially this is achieved by studying two value functions by automatically assigning interactions to upgrade each of the two value functions, resulting in two weight groups, $\theta$ and $\theta$.'
    \item  One collection of weights is used with each update to decide the greedy policy (actions), and the other to calculate its value.
    \item The resulting network is known as Double DQN.
\end{enumerate}

\begin{figure}[H]
    \centering
    \includegraphics[scale=0.2]{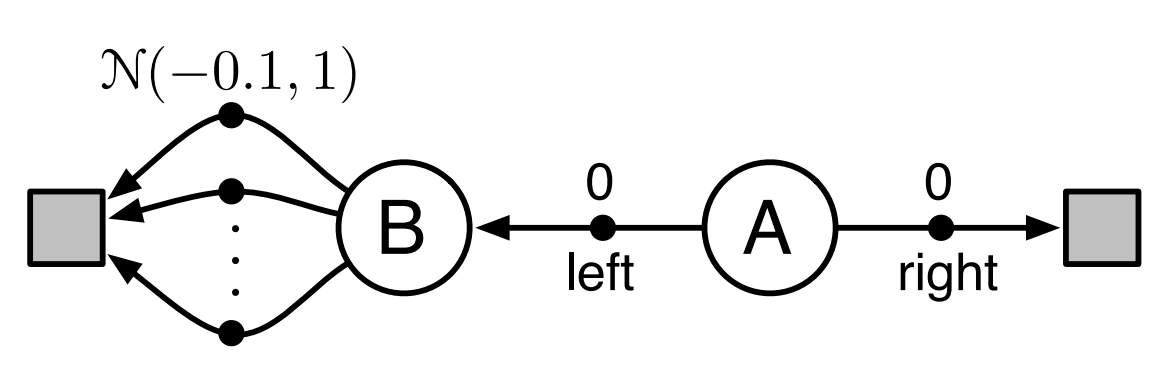}
    \caption{Working of a DQN}
    \label{fig:DQN}
\end{figure}

\subsubsection*{Results}
The testbed consists of Atari 2600 games, using the learning experience of the Arcade.
 \begin{itemize}
    \item DQN is seen as reliably and often greatly over-optimistic about the new greedy policy's worth.
    \item However, DQN had good results in a few games, even after being overoptimistic. 
    \item The agents were often evaluated at various starting points, some provided by humans in the hope of obtaining improved outcomes, and it succeeded, allowing several game results to improve.
\end{itemize}

\subsubsection*{Shortcomings}
An attention-based model could have been attempted in combination with Double DQN to achieve better results.

\subsection{Policy Gradient Method}

\subsubsection*{Introduction}
Method approximation is important for reinforcement learning, but the traditional approach to approximating a value function and deciding a strategy from it has proved logically intractable to date. In this article \cite{ref2}, we discuss an alternate approach in which the policy is directly defined by its own function approximator, independent of the value function, and is modified with respect to the policy parameters according to the gradient of the predicted incentives.

\subsubsection*{Method}
\begin{enumerate}
    \item The value-function strategy is directed towards deterministic measures, while the optimum policy is always stochastic, with complex probabilities choosing various actions. Moreover, an unexpectedly minor difference in an action's expected value will cause it to be picked, or not. These problems were established as a key obstacle to creating algorithm convergence guarantees for the ones following value - function approach.
    \item Hence one uses stochastic policy estimation instead of approximation of a value method . That is used to measure a deterministic approach.
    \item Unlike the approach to value-function,slight changes in policy and distribution of state-visitation corresponds to only tiny changes in parameters.
    \item Hence it proved that generally a policy method with function approximation using differentiation converges to a locally optimal policy.  
\end{enumerate}

\subsubsection*{Results}
\begin{itemize}
    \item The theorem deduced in the research paper assures that the parameters for policy update in the direction of the gradient and hence converge to a local optimum.
    \item Also, it proves the relations between estimated functions and optimal policies.
\end{itemize}

\subsection{High-Frequency Trading}

\subsubsection*{Introduction}
High frequency trading has steadily gained a foothold in capital markets in recent years. Algorithmic trading comprises of about 66 percent of trades in USA and EU. This research paper \cite{ref3} explains the benefits and shortcomings of HFT and implementation details, as well as future scope.

\subsubsection*{Hierarchy of terms}
\begin{enumerate}
    \item Using electronic trading one can transfer requests remotely, and one does not need to depend on any methods involving communication over media or in person. As the inclusion of machines in trading world is increasing, so is the usage of the term. 
    \item In algorithmic trading , the process is automated and there is strict obedience to a specified collection of rules so that execution process can be streamlined. The alogrithms determines various factors involving the market(timing, price,amount, path etc) and keeps on monitoring the market for fluctuations.
    \item It is a subset of algorithmic trading.The orders are done in bulk and very fast, in terms of milliseconds. Also the order size is fairly small. Using the power of faster computation speeds and massive analysis potential of computers, new trade opportunities are opened which might be there for milliseconds and hence blind to the human eye.
\end{enumerate}

\subsubsection*{Strategies}
HFT is not a strategy in itself but incorporates many methods for implementing specific strategies , so that the benefit from them can be maximised. Many short term inefficiencies can be detected.

\begin{enumerate}
    \item First strategy is to provide liquidity and this works like the general market but there is no market making obligation involved formally in the case of HFT. Basically there is a higher bid and ask differential used to profit.
    \item Second strategy focuses at exploiting imbalances in arbitrage securities by the trading algorithm.
    \item To find out larger orders in finding liquidity, one method used is to send out smaller orders and the result is used in predicting the presence of large orders. If the order is completed quickly it confirms the result.
\end{enumerate}

\subsubsection*{Impact Analysis}
Using HFT along with above mentioned strategies has helped the market by having reduced spreads, balanced prices across markets, and more liquidity. However, some issues still remain:
\begin{itemize}
    \item HFTs have no obligations in making the market.
    \item Since the size of quotes and orders is very small there is little contribution to depth of the market. 
    \item There can be false pricing as the orders are available for such short durations that they can be cancelled after the liquidity is realised. 
\end{itemize}

\subsubsection*{Conclusion}
To the point that they aren't still, subjecting HFTs to prudential and operational standards and full regulatory supervision by a competent court could enable the problems to disappear. Even so, HFT has shown that extremely high-speed and powerful computational and algorithmic computer programs are of vital importance for order generation, routing and execution and hence are important for success of the traders.

\section{Proposed System}

\subsection{Introduction}
This section exhibits a deep reinforcement learning framework to device an algorithmic trading model. The framework consists of an environment and an agent. The interaction between them would devise the algorithm.
\subsubsection*{RL}
RL stands for Reinforcement learning. It is a branch of machine learning that works differently compared to the other paradigms. The environments involved are dynamic and complex. The agent tries to maximize the total reward by reinforcement i.e. a reward for an action that leads to maximization of reward and a penalty for an action that reduces the reward.

Hence, there is no requirement for training input-output pairs. Instead, it learns as it interacts with the environment, gradually improving accuracy. RL has proved to be exceptionally useful in use cases such as games and autonomous driving and has potential for more.
\subsubsection*{DRL}
DRL stands for Deep Reinforcement Learning. It utilized the huge dataset space encompassed by neural networks to refine the actions taken by the Reinforcement Learning agent to produce a more optimized output.
 
One can say that a typical RL algorithm has to train on immediate actions having an impact on rewards which are way in the future. However, in certain environments, the action space is very large and a simple lookup table is not optimal to use. Thus, neural networks approximate the functions which render the agent and train the neural network on the action space, optimizing the reward.

Use cases can vary, for example, a convolutional neural net can be used in case of games.

\subsubsection*{DQN}
DQN stands for Deep Q-Learning. One can consider Q-Learning as an algorithm of the creation of the dataset from where the agent builds the policy called a Q-Table. Deep Q-Learning uses a neural network as a Q-Table to build the function approximation. All the experiences are stored in memory. Q-Network decides the best output for maximizing return. A loss function deals with the MSE in the predicted and target value.

Hence Q-Learning algorithm utilizes a neural net to implement a DRL for optimizing the steps for maximum reward.

\subsection{The Data}
The price of financial assets changes over time. Financial time series analysis is the tool used to extract nifty features from the assets. Determining the mathematical parameters such as mean, moving averages,  standard deviation, correlation, covariance, autocorrelation, and convolution can help in obtaining peculiar insights into the data. As our primary motive is to devise a highly lucrative financial portfolio, we performed a few statistical tests from Portfolio management theory to finalize the following assets

The data for 28 assets in the portfolio was extracted from Yahoo finance beginning from Jan 1st, 2010 for stocks and beginning from Jan 1st, 2016 for cryptocurrencies. The time-series data was fed into a Pandas Dataframe for preprocessing. All raw time-series are a combination of these four attributes trend, seasonality, noise, and autocorrelation. The trend is preprocessed with the Detrend functionality of the Signal module available in Python. The seasonality and autocorrelation attributes are exploited to recognize patterns in the time-series. Noise is quantified as the risk associated with the time-series data. The time-series data is collected each day with parameters closed price, open price, highest price, lowest price, volume, and adjusted close price. From various parameters fetched we are only considering the adjusted closed prices. The data points with NA values must be filtered out to avoid further complex computations. 

\subsection{The Environment}
An ideal environment would provide the agent with all the available assets in the financial markets. The total financial information available in the market is quite complicated to quantify and process for the algorithm. Nonetheless, all the information can be assumed to be reflected in the prices on stock markets. So our portfolio consisting of 28 equity assets are fetched through the Yahoo finance and preprocessed. The environment is the medium through which the agent gets the state and decides to act based on exploring or exploitation of Q tables. The environment takes the state and action of the agent to provide a reward. The reward is dispensed at each time-step in an episode. The environment class consists of various functionalities:-

\subsubsection*{Get\_state}
 In reinforcement learning, a state provides us with all the information we need about the environment, and the agent acts accordingly. As our episodes span over eight years of financial time-series we define a novel state for each time-stamp. The first 28 elements in the state tensor consist of pre-processed asset prices from the Pandas Dataframe. The next 28 elements are ten-day moving averages of the respective asset prices. The value ten here is a hyperparameter which is dependent on the choice of aggregated assets in the portfolio. It also varies upon the choice of time-steps. Correlation between two assets quantifies the diversification of risk involved. So the next 784(28*28) elements include a  correlation matrix constructed on the previous ten days of asset prices. Again ten is a hyperparameter depending on the choice of different assert combinations and choice of time-steps. Empirically the value ten gave us better results. 

\subsubsection*{Reset\_state}
Reinforcement learning is a process demanding to run episodes an extensive amount of times. Every time an episode begins the environment needs to be reset to avoid divergence. The time-stamp also needs to be reset.

\subsubsection*{Get\_reward}
The reward function takes into account the current state and the agent's action to reward him accordingly. The choice of reward function is a hyperparameter in itself. The convergence of the model depends upon the preference of the reward function. We choose our reward function from the Portfolio management theory. As our asset data is extracted for each day the daily returns are our choice of the reward function. The daily return on financial assets is the daily percentage change in the price of assets. The reward for each asset is independently fetched to the Q-tables. The cumulative reward on the entire portfolio is the sum of all the rewards.

\subsection{The Agent}

The agent in reinforcement learning is to take action out of the action-space. The agent examines the current state provided by the environment and executes the proper action. In the earlier episodes, the agent needs to explore the action-space with the aid of random function. Then in the later episodes, the explored action-space can be exploited through a neural network.

\subsubsection*{Act}
Conventionally the task of the agent was to buy, hold, or sell an asset in the portfolio and was rewarded proportionately. The actions were then consigned weights. We intend to eliminate the middle step. So, the task of our agent is to assign weights to assets in the portfolio to earn higher returns. A rudimentary agent could either go long(buy at a low price and then sell at a high price) or short(sell at a high price and then buy at a low price).

\begin{enumerate}
    \item The Long:- The exploration process assigns weights through a uniform random function in range 0 to 1 in a vector. The vector then is divided by the vector sum to make the sum of elements one. 
    \item The Short and the long:- The exploration process assigns weights through a uniform random function in the range -1 and 1 in a vector. The vector is divided by the absolute vector sum to make the sum of elements one.
\end{enumerate}

The initial investment amount is distributed depending upon the weights assigned. As the weights alter each episode the agent must perform financial transactions internally to allocate the investment proportionately. The environment rewards the agent at the end of each episode which is cumulated with the initial investment and reinvested. 

\subsubsection*{Replay Buffer}
The exploration process is just a random weight assignment, but the task is to learn the optimal policy. The agent only has the information about the latest actions and the environment has the latest rewards. The Deep Q network trained on the latest experience is just a random agent. So the replay buffer provides a means to record the experience. This experience is a set of the previous state, current actions, current state, and the reward. The experience is pushed into a fixed-sized buffer of 32 with first in first out strategy. The size of the buffer is another hyperparameter. The deep Q network is trained in a mini-batch of buffer size with the state as the input and the actions as the outputs. As every batch is processed we move in the direction of optimal policy. 

\subsection{The Model}
In our study, there are 1000 episodes to train the Deep Q network. At the beginning of each episode, the state and the time-stamp must be reset through the environment. For each step in the episode, time-stamp is incremented by a unit. The environment fetches the state at each step for the agent to take action. The agent then decides to explore or exploit the action-space. A random function is used for exploration and a neural network for exploitation. A set is formed with the previous state, the actions, and the current state. The environment determines the reward for the current policy with this set. The reward is a weight vector of size 28. The reward is then attached to the above set and stored in the replay buffer. The replay buffer follows the property of first in first out. Once the buffer is filled, the sets are passed as arguments to the expReplay function. Here, the Q-tables are updated with the rewards for every 28 elements. The state and Q-values are passed to the neural network model where the loss function is computed. It is the difference between the output of the model and the Q-values. So the neurons in the network are re-computed to move in the direction of the optimal policy for every mini-batch of size 32.
The current state is fetched as the input to the neural network. There are 28 output layers each designed to assign a weight to the assets. The output of each layer is the Q-value which is processed to compute the respective weights. The sum of these weights is always one.

\begin{figure}[H]
    \centering
    \includegraphics[scale=0.1]{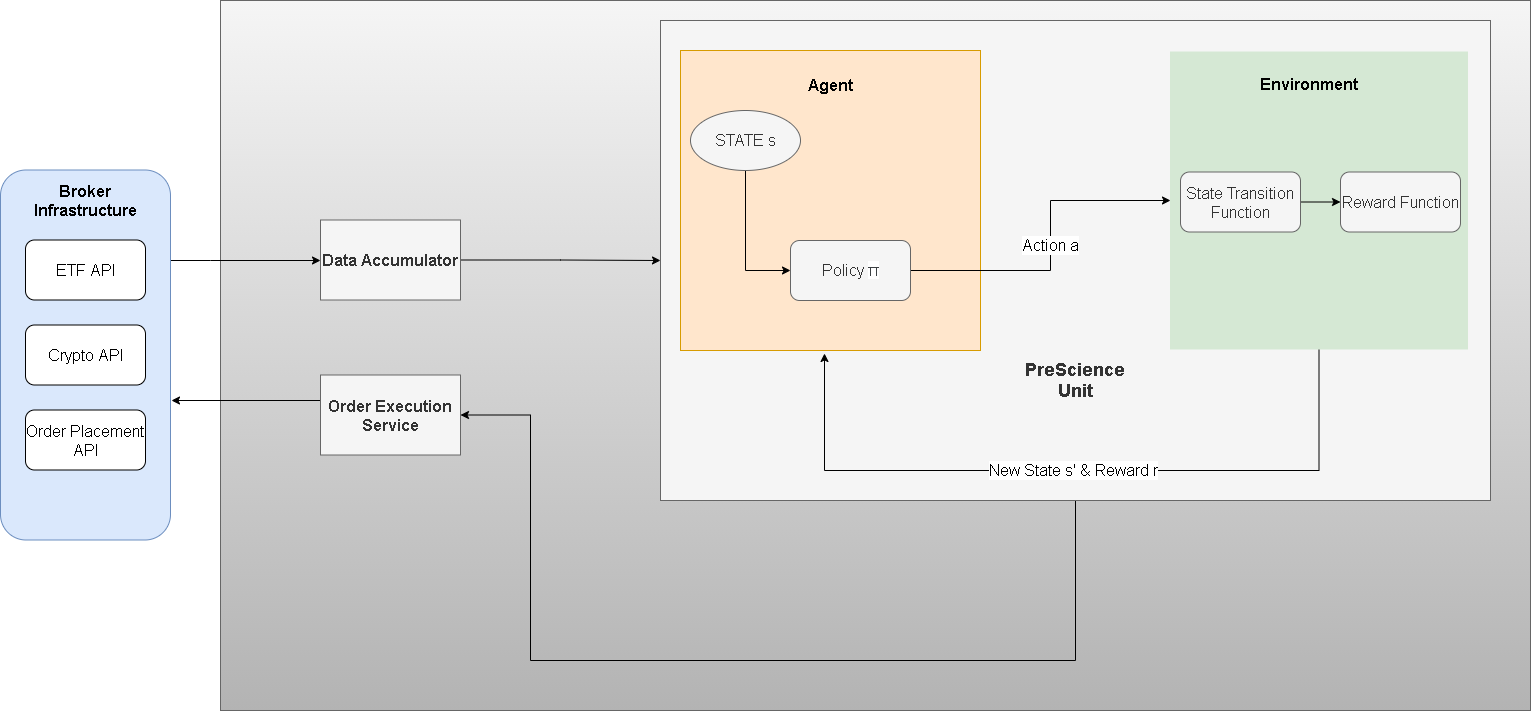}
    \caption{RL Flow Chart}
    \label{fig:galaxy}
\end{figure}

\section{Analysis and Results}
\subsection{Introduction}
In this section, we will compare the efficiency of our RL model as compared to pre-existing methods(performance metrics). The data set comprises of cryptocurrencies and ETFs.
\subsection{Performance Metrics}
Following performance parameters are used to express the returns on the input given to various approaches :-
\subsubsection*{Mean Returns}
 As the name suggests, it is the mean/ average value of all the returns provided by a portfolio/ stock.
\subsubsection*{Volatility}
  Volatility is a measure of the risk involved in a stock. By definition, it is the range over which the returns are spread. It can be measured as standard deviation or variance.
\subsubsection*{Sharpe Ratio}
Sharpe Ratio is also a measure of the riskiness of a stock. Mathematically, it is the average excess return per unit of volatility.
\subsubsection*{Alpha}
Alpha is the excess return on an investment, however it is relative to a benchmark index.
\subsubsection*{Beta}
Beta is volatility of a stock in a complete market. Hence it is the measure risk associated with a stock in a diverse market.

\subsection{Results}
\subsubsection*{Min Variance}
\begin{figure}[H]
    \centering
    \includegraphics[scale=0.30]{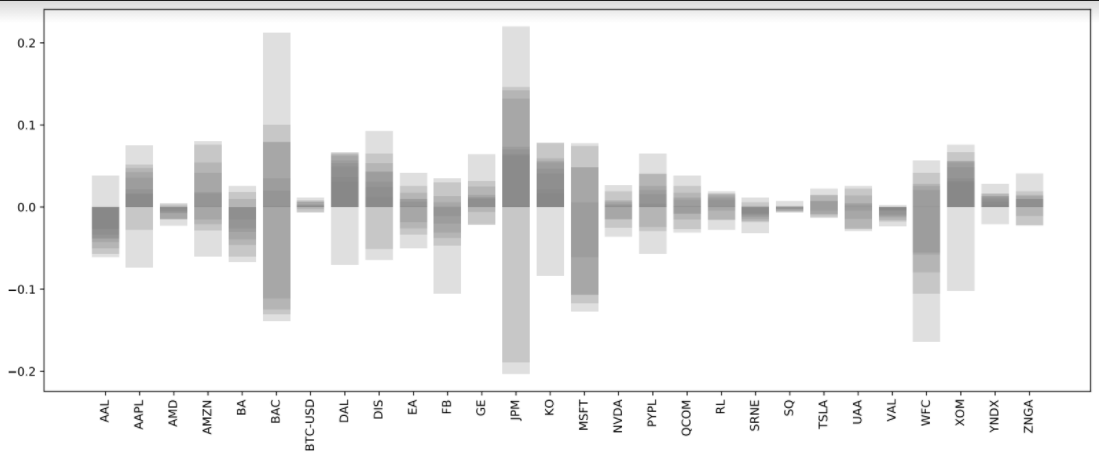}
    \caption{Performance of Min Variance Model}
    \label{fig:min_variance}
\end{figure}

Performance of Min Variance for the portfolio is shown in Figure \ref{fig:min_variance}

\subsubsection*{Max Returns}
\begin{figure}[H]
    \centering
    \includegraphics[scale=0.30]{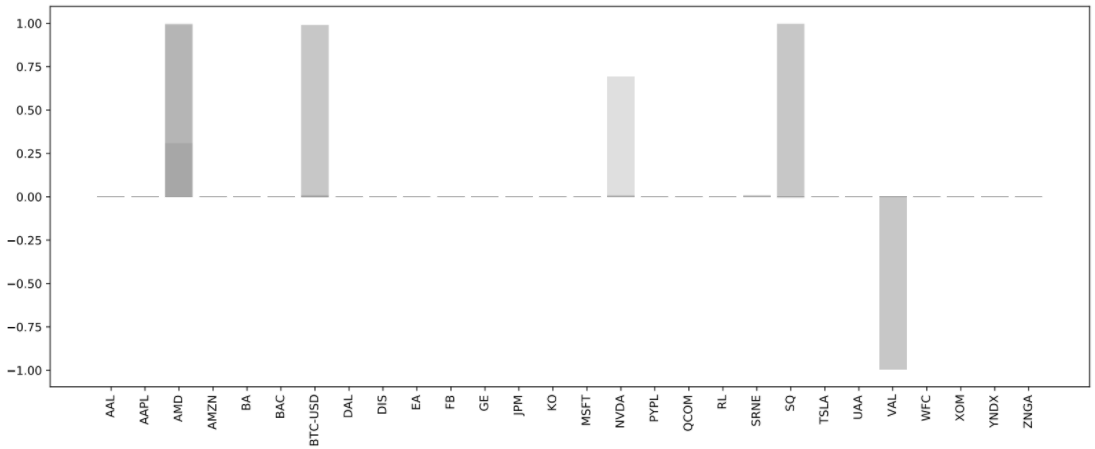}
    \caption{Performance of Max Returns Model}
    \label{fig:max_returns}
\end{figure}

Performance of Max Returns for the portfolio is shown in Figure \ref{fig:max_returns}

\subsubsection*{Auto-Encoder}
\begin{figure}[H]
    \centering
    \includegraphics[scale=0.30]{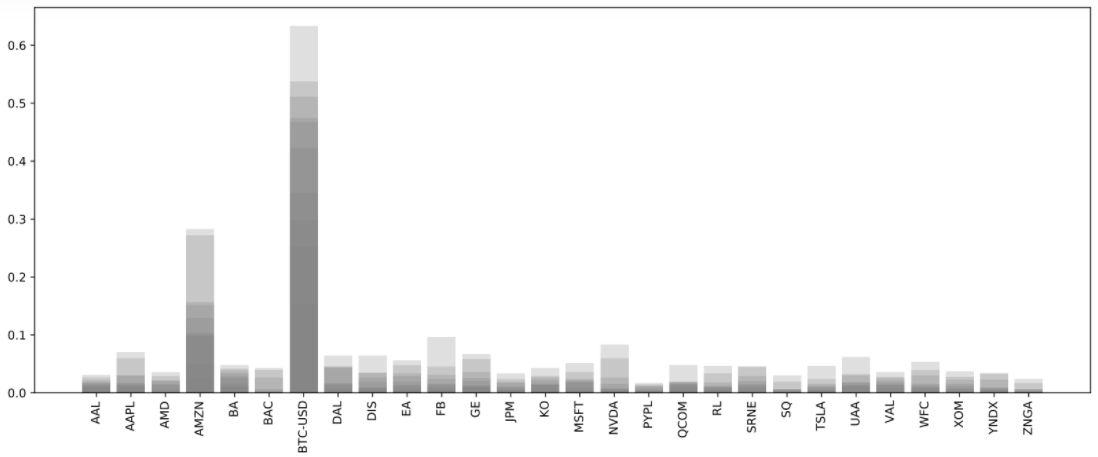}
    \caption{Performance of auto-encoder Model}
    \label{fig:auto_encoder}
\end{figure}

Performance of Auto-Encoder for the portfolio is shown in Figure \ref{fig:auto_encoder}

\subsubsection*{Deep Reinforcement Learning}
\begin{figure}[H]
    \centering
    \includegraphics[scale=0.30]{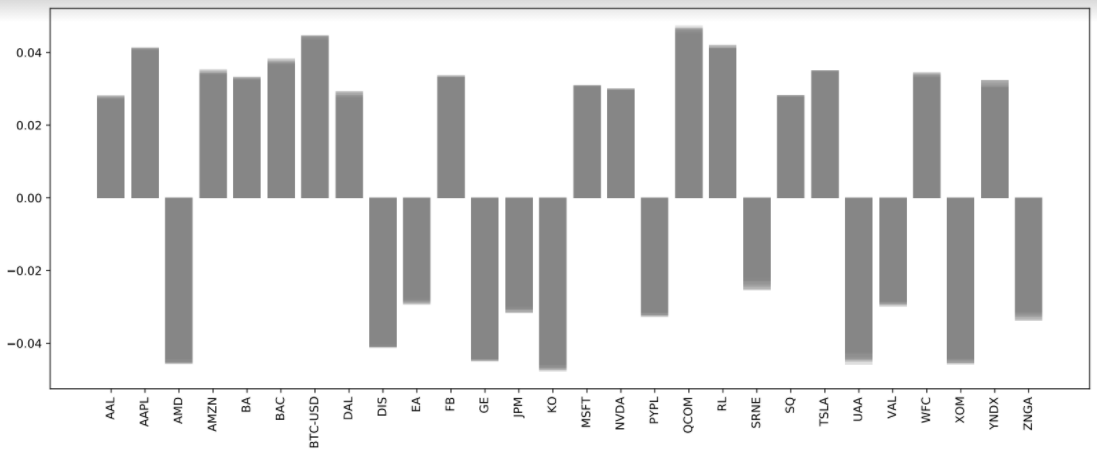}
    \caption{Performance of DRL Model}
    \label{fig:drl}
\end{figure}

Performance of Deep Reinforcement Learning for the portfolio is shown in Figure \ref{fig:drl}

\subsection{Analysis}
\begin{figure}[H]
    \centering
    \includegraphics[scale=0.3]{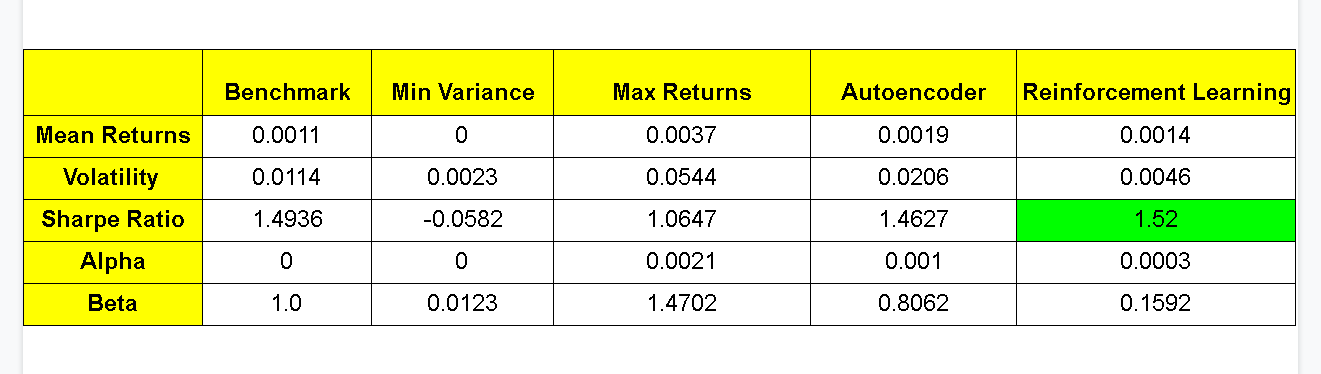}
    \caption{Comparison of RL with conventional portfolio management agents}
    \label{fig:rl_conventional}
\end{figure}

Now, to compare the results of our agent, the results shown in Figure \ref{fig:rl_conventional} are obtained

Hence, the proposed model gave the highest Sharpe Ratio.

\section{Conclusion}
This work presented a deep reinforcement network based approach to allocate portfolio funds. The proposed model is aimed at maximizing returns with minimum risk exposure. The use of LSTM in the field of stock data modeling has been an active area of research and can be accustomed to the proposed model for better results in future work. There have been developments in Reinforcement Learning architectures like A3C, PPO, etc making the model more versatile for market shocks. This work provides an empirical conclusion that portfolio fund allocation through deep reinforcement weight learning could outperform in risk-adjusted returns conventional portfolio management agents.

\end{document}